\documentclass[runningheads]{llncs}

\usepackage{graphicx}
\usepackage{booktabs}
\usepackage{amsmath}
\usepackage{amssymb}
\usepackage[dvipsnames]{xcolor}
\usepackage{hyperref}

\begin{document}

\title{Open-Domain Conversational Search Assistant with Transformers}

\subtitle{A Preprint}

\titlerunning{Open-Domain Conversational Search Assistant w/ Transformers - Preprint}

\author{Rafael Ferreira \and Mariana Leite \and
\\
David Semedo \and
Joao Magalhaes}
\authorrunning{R. Ferreira et al.}
%
\institute{NOVA LINCS, School of Science \& Technology,\\ NOVA School of Science and Technology, Portugal\\
\email{\{rah.ferreira, me.leite\}@campus.fct.unl.pt},
\email{\{df.semedo, jmag\}@fct.unl.pt}}

\maketitle              
\begin{abstract}
Open-domain conversational search assistants aim at answering user questions about open topics in a conversational manner. 
In this paper we show how the Transformer architecture~\cite{vaswani2017attention} achieves state-of-the-art results in key IR tasks, leveraging the creation of conversational assistants that engage in open-domain conversational search with single, yet informative, answers. 
In particular, we propose an open-domain abstractive conversational search agent pipeline to address two major challenges: first, conversation context-aware search and second, abstractive search-answers generation. 
To address the first challenge, the conversation context is modeled with a query rewriting method that unfolds the context of the conversation up to a specific moment to search for the correct answers. These answers are then passed to a Transformer-based re-ranker to further improve retrieval performance. 
The second challenge, is tackled with recent Abstractive Transformer architectures to generate a digest of the top most relevant passages. 
Experiments show that Transformers deliver a solid performance across all tasks in conversational search, outperforming the best TREC CAsT 2019 baseline.

\end{abstract}

\section{Introduction}
Conversational search systems are an emerging research topic, and the natural evolution of the traditional search paradigm, allowing for a more natural interaction between users and search systems. 
Building intelligent systems able to establish and develop meaningful conversations is one of the key goals of AI and the ultimate goal of natural language research~\cite{wizardofwikipedia}. 
The interactions between a user and conversational systems have been studied in ~\cite{evaluation_conversational_assistants}, which showed that users are willing to utilise conversational assistants as long as their needs are met with success.
However, conversational search assistants still put a considerable burden on users that have to go through a list of documents, or passages, to find the information they need.

We depart from this document-based approach to conversational search, and propose an open-domain abstractive conversational assistant that is aware of the context of the conversation to generate a single and informative search-answer.
We argue that by doing so, we can capture in one single and short answer the information contained on several relevant documents. 
Moreover, we show that Transformer architectures~\cite{vaswani2017attention} outperform the state-of-the-art results across all the steps of the conversational system pipeline.
Hence, the core contributions of this paper are twofold: first, we show that one can tightly integrate different Transformers to deliver an end-to-end conversational search pipeline with state-of-the-art results; second, abstractive answer generation can effectively compress the information of several retrieved passages into a short answer.
These contributions are rooted in the groundbreaking architecture of the Transformer~\cite{vaswani2017attention} that leverages attention mechanisms to model complex interactions between sequence data. 
In particular, we explore Transformer's advantages to: (a) capture complex relations between conversation turns to rewrite a query in the middle of a conversation; (b) to look into the interactions between words in a conversation query and a candidate passage; and (c) to compress multiple retrieved passages into one single, yet informative, search-answer. 
The final result, is a complete conversational search assistant leveraged by the Transformer architecture.

In the following section, we discuss the related work. In section 3 we detail the Transformer-based conversational search pipeline: the conversational query rewriting, the re-ranker, and abstractive answer generation. Evaluation is performed in Section 4 and Section 5 presents the key takeaway messages.

\section{Related Work}
Open-domain conversational search systems must account for the dialog context to provide a relevant passage. While research on interactive search systems has started long ago~\cite{belkin1980anomalous,croft_newapproach_i3r,oddy_information_1977}, the recent interest in having intelligent conversation assistants (e.g. Alexa, SIRI), has re-ignited this research field. Recent models~\cite{wizardofwikipedia,t5conversational,open-retrieval-qa-sigir2020,limited_supervision_query_rewrite} leverage large open-domain collections (e.g. Wikipedia) to learn rich language-models using self-supervised neural networks. The applicability of these models in conversational search is twofold: grasping the dialog context and passage re-ranking. Recently, the TREC CAsT (Conversational Assistant Track)~\cite{castoverview} task introduced a multi-turn passage retrieval dataset, enabling the development and evaluation of such models.

Conversational context-aware search models need to (a) keep track of the dialog context, and (b) select the most relevant passage. 
To address (a), one approach is to perform query rewriting to obtain context-independent queries.~\cite{canYouUnpackIt} observed that manually rewritten queries from QuAC~\cite{quac_dataset} had enough context to be independently understandable. To automate the process, a sequence-to-sequence (seq2seq) model with attention and a copy mechanism was proposed. The model is given as input a sequence with the full conversation history and the query to be rewritten. In~\cite{limited_supervision_query_rewrite}, a BERT model~\cite{bertOriginal} is given as input a sequence of all terms of the current and previous queries, and is then fine-tuned on a binary term classification task.
Also using both the query and conversation history, in~\cite{t5conversational}, a pre-trained T5 model~\cite{t5} is fine-tuned on CANARD~\cite{canYouUnpackIt} to construct the context-independent query, and achieved state-of-the-art performance on the query-rewriting task. Task (b) is commonly addressed through re-ranking.
Large pre-trained Transformer models, such as BERT~\cite{bertOriginal}, RoBERTa~\cite{roberta}, and XLNet~\cite{yang2019xlnet}, have been widely adopted for re-ranking due to their generalisation capabilities.
Examples of this are present in~\cite{han2020learningtorank,passagererankingbert,nogueira2019multistage}, where a Transformer-based model is fine-tuned on the question-answering relevance classification task.

Given the dialogue context, the agent must generate a natural language response. 
In chit-chat dialogue generation, most approaches use an encoder-decoder neural architecture that first encodes utterances and then the decoder generates a response~\cite{reinforcement_dialog_generation,adversarial_dialogue_generation,retrieval_based_generation_1,memory_augmented_conversational_response,retrieval_based_generation_2}. In~\cite{reinforcement_dialog_generation} and~\cite{adversarial_dialogue_generation}, reinforcement learning is used to overcome uninformative and general responses of standard seq2seq models.
Another alternative is retrieval-based dialogue generation, in which the generator takes as input retrieved candidate documents to improve the comprehensiveness of the generated answer~\cite{retrieval_based_generation_1,retrieval_based_generation_2}. These approaches require a large dataset with annotated dialogues, which is not feasible in our scenario. Alternatively, Transformer models have shown to be highly effective generative language models~\cite{bart,t5,pegasus}. While both T5~\cite{t5} and BART~\cite{bart} are general language models, PEGAGUS~\cite{pegasus} focuses on abstractive summarisation, and obtained state-of-the-art results on 12 summarisation tasks. 

\section{Transformers-based Conversational Search Assistant}
\begin{figure}[t]
  \centering
    {\includegraphics[width=1.0\linewidth, trim=1pt 1pt 1pt 1pt, clip]{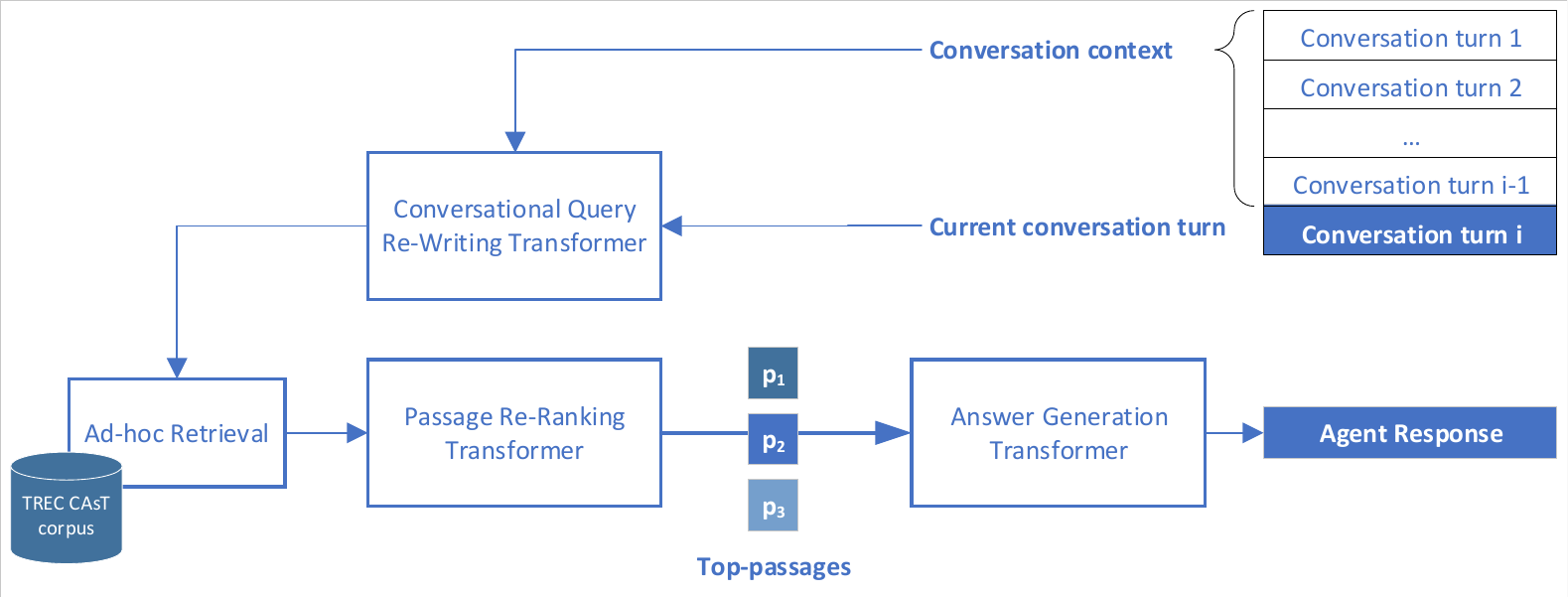}}
  \caption{The proposed Transformer-based conversational search assistant.}
  \label{fig:conversationalarchitecture}
\end{figure}

In this section we formulate the open-domain conversational search task and describe the conversational assistant retrieval and answer generation components.
The conversational search task is formally defined by a sequence of natural language conversational turns for a topic $T$, with queries $q$. For each conversation turn $T=\{q_1,...q_i,...q_n\}$, the conversational search task is to find relevant passages $p_k$ for each query $q_i$, satisfying the user’s information need for that turn according to the conversational context. 
The proposed approach uses a four-stage architecture: (a) context tracking, (b) retrieval, (c) re-ranking, and (d) answer generation. 
An overview of the system's architecture can be seen in Figure~\ref{fig:conversationalarchitecture} which we will detail in the following sections.

\subsection{Conversational Query Rewriting Transformer}
Due to the evolving nature of a conversational session, the current query may not include all the information needed to retrieve the answer that the user is looking for. This challenge is illustrated in the conversation presented in Table~\ref{table:example_conversation_lucca}: in conversation turn 2, the system needs to understand that ``its"\ refers to ``Lucca's"\ (explicit coreference) and in turn 3, where the important monuments should be focused in Lucca, although there is no direct evidence (implicit coreference), which makes the task even more challenging. 
We tackle this challenge by rewriting queries, using previous turns, making the current query context-independent.
\begin{table}[htbp]
\centering
\caption{Conversation example about a specific topic, in this case the city of Lucca.}
\label{table:example_conversation_lucca}
\begin{tabular*}{\textwidth}{cll}
\hline
\textbf{Turn} & \textbf{Conversational Query}                & \textbf{Context-independent Query} \\ \hline
1 & How is the climate in \underline{Lucca}? & How is the climate in \underline{Lucca}? \\
2 & Tell me about \underline{its} origins.    & Tell me about \underline{Lucca's} origins. \\
3 & What monuments should I visit? \ & What monuments should I visit \underline{in Lucca}? \\ \hline
\end{tabular*}%
\end{table}

To perform the query rewriting task, we need a model capable of performing coreference resolution and include context from previous turns. The Text-to-text Transfer Transformer (T5)~\cite{t5} can be fine-tuned to reformulate conversational queries~\cite{t5conversational} by providing as input the sequence of conversational queries and passages, and as target, the rewritten query.
The training input sequence is constructed as: 
\begin{equation}
\label{t5_input}
    ``q_i \ [CTX] \ q_1 \ p_1 \ [TURN] \ q_2 \ p_2 \  [TURN]\  \ldots \ [TURN] \ q_{i-1} \ p_{i-1}",    
\end{equation}
\noindent
where $i$ is the current turn, $q$ is a query, $p_k$ is a passage retrieved from the index by the retrieval model, and $[CTX]$ and $[TURN]$ are special tokens. $[CTX]$ is used to separate the current query from the context (previous queries and passages) and $[TURN]$ is used to separate the historical turns (query-passage pair).

\subsection{Passage Re-ranking Transformer}
With the new pre-trained neural language models, such as BERT~\cite{bertOriginal} and others~\cite{roberta,yang2019xlnet}, it is possible to generate contextual embeddings for a sentence and each of its tokens. These embeddings can be used as input to a model to perform passage re-ranking~\cite{passagererankingbert,nogueira2019multistage}. This re-ranking step allows going beyond term matching, as the model has some understanding of both individual terms semantics as well as their interactions between queries and passages. As such, it is able to judge more thoroughly if a passage is relevant to a query. 

Following this rationale, we tackle the passage re-ranking task with a BERT model~\cite{bertOriginal}, fine-tuned on the passage ranking task~\cite{passagererankingbert}, through a binary relevance classification task, where positive examples are relevant passages, and negative examples are non-relevant passages. 
To obtain the embedding of the query $q$, and passage $p$, 
a sequence with $N$ tokens is given as input to BERT:
\begin{equation}
\label{eq:embedding_bert}
    emb = BERT(``[CLS]\ q \ [SEP] \ p"),
\end{equation}
\noindent
where $emb \in \mathbb{R}^{N \times H}$ ($H$ is BERT embedding's size) is the embeddings matrix of all tokens, and \textit{[CLS]} and \textit{[SEP]} are special tokens in BERT's vocabulary, representing the classification and separation tokens, respectively.
From $emb$ we extract the embedding of the first token, which corresponds to the embedding of the  \textit{[CLS]} token, $emb_{[CLS]} \in \mathbb{R}^{H}$. This embedding is then used as input to a single layer feed-forward neural network (FFNN), followed by a \textit{softmax}, to obtain the probability of the passage being relevant to the query: 
\begin{equation}
\label{eq:bert_passage_raking}
    P(p|q)=softmax( \text{FFNN}(emb_{[CLS]}) ).
\end{equation}
\noindent
With $P(p|q)$ calculated for each passage $p$ given a query $q$, the final rank is obtained by re-ranking according to the probability of being relevant.

\subsection{Abstractive Search-Answer Generation Transformer}
Having identified a set of candidate passages according to the scores given by the re-ranker model (equation~\ref{eq:bert_passage_raking}), the goal is to generate a natural language response that combines the information comprised in each of the passages. To address this, we follow an abstractive summarisation approach, which unlike extractive summarisation that just selects existing sentences, it can portray both reading comprehension and writing abilities, thus allowing the generation of a concise and comprehensive digest of multiple input passages. 

The Transformer~\cite{vaswani2017attention} architecture has proved to be highly effective at modelling large dependency windows of textual sequences. Text-to-text approaches \cite{bart,t5,pegasus}, trained over large and comprehensive collections, become effective at \textit{understanding} different topics and retaining language regularities useful for several language tasks. 
Thus, to generate the agent's response using a transformer model, we give as input the  following sequence: 
\begin{equation}
\label{summarization_input}
    ``p_1\  p_2\  \ldots\  p_N",    
\end{equation}
\noindent
where each $p_k$ corresponds to one of the top-N candidate passages. With this strategy, we implicitly bias the answer generation by asking the model to summarise the passages that are deemed as more relevant according to the retrieval component. 

The implicit bias of the top passages is crucial to steer the Transformer response generation. The sequence of passages of equation~\ref{summarization_input} is given as input to the Transformer, which will then attend to the different passages.
As the multi-head attention layers look across the different passages, redundant parts will be merged, while the remaining information will be summarised, leading to a concise but comprehensive answer.
The following Transformer models were considered for the task of abstractive summarisation:

\begin{itemize}
    \item \textbf{Text-to-Text Transfer Transformer (T5)~\cite{t5}} is a text-to-text model based on the encoder-decoder Transformer architecture, pre-trained on the large C4 corpus, which was derived from Common Crawl\footnote{\url{https://commoncrawl.org/}.}. A masked language modelling objective is used, where the model is trained to predict corrupted randomly sampled tokens, of varying sizes.
    \item \textbf{BART~\cite{bart}} is a denoising autoencoder, that combines Bidirectional and Auto-Regressive Transformers.  Pre-training consists of corrupting text with an arbitrary noising function and learning an autoencoder to reconstruct the original text. The best performing noise functions were text infilling (using single mask tokens to mask random sampled spans of text), and sentence shuffling (changing the order of sentences in passages). 
    \item \textbf{PEGASUS~\cite{pegasus}} specialises on the abstractive summarisation task. Multiple important sentences are masked and used as targets, i.e., the model is trained to generated each omitted sentence as output. As in T5, this model is not trained to reconstruct sequences.
\end{itemize}

\section{Evaluation}

\subsection{Datasets and Protocol}
\subsubsection{CANARD Dataset~\cite{canYouUnpackIt}.}
This dataset was used to train and evaluate the query rewriting method. It was created by manually rewriting the queries in QuAC~\cite{quac_dataset} to form non-conversational queries. The training, development, and test sets have 31.538, 3.418, and 5.571, query-rewrites respectively.

\subsubsection{TREC CAsT Dataset~\cite{trecCast}.} This dataset was used to evaluate both the conversational search and answer generation components. 
There are 50 evaluation topics, each with about 10 turns. Of those in total, 20 conversational topics were labelled on average until turn depth 8 using a graded relevance that ranges from 0 (not relevant) to 4 (highly relevant).
The passage collection is composed by MS MARCO~\cite{marcoDataset}, TREC CAR~\cite{treccar}, and WaPo~\cite{washingtonDataset} datasets, which creates a complete pool of close to 47 million passages.  

\subsubsection{Experimental Protocols.}
To analyse query rewriting performance, we used the BLEU-4 score~\cite{bleu} between the model’s output and the queries rewritten by humans, on the CANARD dataset.

In the passage retrieval experiment, we used the TREC CAsT setup and the official metrics, nDCG@3 (normalised Discounted Cumulative Gain at 3), MAP (Mean Average Precision), and MRR (Mean Reciprocal Rank), along with Recall and P@3 (Precision at 3). 

In the answer generation experiment, we used METEOR and the ROUGE variant ROUGE-L. 
For each query in TREC CAsT, we use as reference passages, all the passages with a relevance judgement of 3 and 4. Hence, the goal is to generate answers that cover, as much as possible, the information contained in all relevant passages, in one concise and summarised answer.

\subsection{Implementation}
\subsubsection{Query Rewriting.}
We fine-tuned the T5~\cite{t5} model according to~\cite{t5conversational} and used the CANARD's training set~\cite{canYouUnpackIt}, providing as input the concatenation of the conversational queries and passages, and as target the rewritten query. In particular, we used the T5-BASE model and trained for 4000 steps, using a maximum input sequence length of 512 tokens, a maximum output sequence length of 64 tokens, a learning rate of 0.0001, and batches of 256 sequences. 

\subsubsection{First-stage Retrieval.}
To index and search, we used the well tuned Anserini framework~\cite{anserini}, in particular, the Python implementation Pyserini\footnote{\url{https://github.com/castorini/pyserini}}. We applied stop word removal, using Lucene’s default list, and stemming using Kstem\footnote{\url{http://lexicalresearch.com/kstem-doc.txt}}. 
We experimented with: BM25~\cite{bm25}, language models with Dirichlet (LMD) and Jelinek-Mercer (LMJM) smoothing~\cite{languagemodelsmoothing} and from our initial analysis, LMD showed better results. This confirms previous knowledge~\cite{languagemodelsmoothing} and matches the shorter queries that we observe in a conversational search scenario. Hence, LMD was the model used in all experiments.

\subsubsection{BERT Passage Re-ranker.}
To perform re-ranking, we used the BERT model implementation from Huggingface~\cite{huggingface}. 
Following the state-of-the-art~\cite{passagererankingbert,nogueira2019multistage}, we used the LARGE version of BERT with a classification layer (feed-forward neural network) on top, that takes as input the query-passage \textit{CLS token} embeddings vector generated by BERT, and classifies the passage as relevant or non-relevant to that query. 
This model was trained following~\cite{passagererankingbert} on the MS MARCO dataset~\cite{marcoDataset}.
In testing, we truncate the concatenation of the query, passage, and separator tokens to a maximum of 512 tokens (the maximum number of tokens for the BERT model). 

\subsubsection{Transformer based answer generation.}
To generate the summarised answers, we employed the T5-BASE, BART-LARGE and PEGASUS models~\cite{huggingface}. The T5-BASE has about 220 million parameters with 12 layers, 768 hidden-state size, 3072 feed-forward hidden-states and 12 heads. BART-LARGE holds about 406 million parameters, with a 12-layer, 1024 hidden state size and 16-head architecture. The PEGASUS model has the biggest number of parameters,  568 million, with 16 layers, 1024 hidden state size and 16-heads.

All models were fine-tuned on the summarising task with the CNN/Daily Mail dataset~\cite{cnndailymail}. To generate the summary, we use 4 beams, restrict the n-grams of size 3 to only occur once, and allow for beam search early stopping when at least 4 sentences are generated. Additionally, we fix the maximum length of the summary to be of the same length of the input given to the models (which corresponds to 3 passages) and vary the minimum length from 20 to 120 words.

\subsection{Results and Discussion}
\subsubsection{Conversation-aware Query Rewriting.}
In Table~\ref{table:t5_bleu_results}, we show the BLEU-4 scores obtained in CANARD's test set and in TREC CAsT's 2019 manually rewritten queries. The rows ``Human" and ``Raw" are from~\cite{canYouUnpackIt}, the row ``T5-BASE" is from~\cite{t5conversational}. The last row corresponds to our implementation. Our results are on par with~\cite{t5conversational}, being lower in the CANARD dataset but higher in TREC CAsT. 
We believe the minor differences in performance between our T5-Base model and the T5-BASE from~\cite{t5conversational} are due to the use of different input sequences, as the exact method of constructing the input is not specified in~\cite{t5conversational}.
\begin{table}[htbp]
\centering
\caption{BLEU-4 scores for the CANARD test set and for TREC CAsT using the manually rewritten queries of the evaluation set.}
\label{table:t5_bleu_results}
\begin{tabular}{lcc}
\toprule
{} & \textbf{CANARD}  & \textbf{TREC CAsT} \\ \midrule
{Human~\cite{canYouUnpackIt}} &  59.92 & {-} \\
{Raw~\cite{canYouUnpackIt}}   & {47.44} & {-}  \\
{T5-BASE~\cite{t5conversational}} & {\textbf{58.08}} & {75.07} \\ 
{Our T5-BASE}      & {56.84} & {\textbf{79.67}}     \\ \bottomrule
\end{tabular}
\end{table}

From the analysis of the BLEU-4 scores and outputs, we can conclude that the model is performing both coreference and context resolution, approximating the queries in a conversational format to context-independent queries. 
Examples of the inputs, targets, and predicted queries, are presented in Table~\ref{table:t5example}. 
In TREC CAsT, the historical utterances do not depend on the responses of the system, so the answer is not provided as input. As we can see, T5 is capable of resolving ambiguous queries by co-reference resolution, as in example 1, but sometimes mistakes similar co-references when multiple are involved, as evidenced in example 2 and in~\cite{t5conversational}, where the model predicts ``throat cancer" instead of ``lung cancer". We can also note that this model is more robust than just coreference resolution, as seen in example 3, where it includes the words ``Bronze Age Collapse", even though there is no explicit mention (implicit coreference).

\begin{table}[t]
\centering
\caption{Example of query rewriting inputs, targets and predictions.}
\label{table:t5example}
\resizebox{\textwidth}{!}{%
\begin{tabular}{lp{0.95\linewidth}p{0.2\linewidth}}
\toprule
\multicolumn{2}{c}{\textbf{CANARD}}\\
\midrule
\textbf{Original Query} &  What was \underline{his} agreement with McMahon? \\
  \textbf{T5 Input Query} &  What was his agreement with McMahon? {[}CTX{]} Superstar Billy Graham. Return to WWWF (1977-1981) {[}TURN{]} Why did he return to the WWWF? An agreement with promoter Vincent J. McMahon Senior. \\
  \textbf{T5 Predicted Query} & What was \underline{Superstar Billy Graham's} agreement with McMahon? \\
  
  \textbf{Target Query} & What was \underline{Billy Graham's} agreement with McMahon? \\
\midrule
\multicolumn{2}{c}{\textbf{TREC CAsT 2019}}\\
\midrule
  \textbf{Original Query} &  What are \underline{its} symptoms? \\
  \textbf{T5 Input Query} &  What are its symptoms? {[}CTX{]} What is throat cancer? {[}TURN{]} Is throat cancer treatable? {[}TURN{]} Tell me about lung cancer. \\
  \textbf{T5 Predicted Query} & What are \underline{throat cancer's} symptoms? \\
  \textbf{Target Query} & What are \underline{lung cancer's} symptoms? \\
\midrule
  \textbf{Original Query} & What are some of the possible causes? \\
  \textbf{T5 Input Query} & What are some of the possible causes? {[}CTX{]} Tell me about the Bronze Age collapse? {[}TURN{]} What is the evidence for the Bronze Age collapse? \\
  \textbf{T5 Predicted Query} & What are some of the possible causes \underline{for the Bronze Age} \underline{collapse}? \\
  \textbf{Target Query} & What are some of the possible causes \underline{of the Bronze Age collapse}? \\
\bottomrule
\end{tabular}%
}
\end{table}

\subsubsection{Transformer-based Passage Search.}
Table~\ref{tab:retrieval_results_simple} shows the results of retrieval on the TREC CAsT dataset. \textit{Original} are the conversational queries (lower-bound), \textit{Manual} is a baseline where the queries were manually rewritten (upper-bound), \textit{T5} is using our query rewriting method, and the other two lines are the results of baselines retrieved from~\cite{castoverview}. \textit{clacBase}~\cite{Clarke2019WaterlooClarkeAT} is a method that uses AllenNLP coreference resolution~\cite{AllenNLP} and a fine-tuned BM25 model with pseudo-relevance feedback, and HistoricalQE~\cite{QueryAnswerExpansionfromConversationHistory} is a method that uses a query expansion algorithm based on session and query words together with a BERT LARGE model for re-ranking. The latter was the best performing method in terms of nDCG@3 in TREC CAsT 2019~\cite{castoverview}.

The first observation that emerges from Table~\ref{tab:retrieval_results_simple} is the clear need for a query rewriting method to maintain the conversational context, evidenced by the low scores on all metrics using the original conversational queries. Rewriting queries (with the \textit{T5} model) outperforms the original conversational queries by a $5-20\%$ margin (nDCG@3), thus showing the effectiveness of this approach.
The second clear observation is again the considerable improvement when Transformers are used for re-ranking. In this case, the improvement is in the $10-15\%$ range over standard retrieval metrics. This is due to the better understanding that the fine-tuned BERT model has of the interactions between the query and passage terms.

Finally, the largest gains emerge when we combine the two Transformers to deliver state-of-the-art results. With the proposed Transformers we outperform the best TREC CAsT 2019 baseline by $3.9\%$ in terms of nDCG@3. We consider that this improvement is mainly due to the use of a better query-rewriting method that allows the retrieval model to retrieve passages given the conversational context, providing the re-ranker with more relevant passages.

\begin{table}[t]
\centering
\caption{Results of retrieval on the TREC CAsT evaluation set.
The HistoricalQE~\cite{QueryAnswerExpansionfromConversationHistory} was the best performing model in TREC CAsT 2019.}
\label{tab:retrieval_results_simple}
\begin{tabular}{llccccc}
\toprule
\textbf{Queries} & \textbf{Re-ranker} & \textbf{Recall} & \textbf{P@3} & \textbf{MAP} & \textbf{MRR} & \textbf{nDCG@3} \\ \midrule
Original             & - & 0.454 & 0.262 & 0.141 & 0.336 & 0.167 \\
Original             & BERT & 0.454 & 0.385 & 0.181 & 0.456 & 0.272 \\ 
T5               & - & 0.697 & 0.474 & 0.251 & 0.597 & 0.322 \\
T5               & BERT & 0.697 & 0.632 & \textbf{0.310} & \textbf{0.739} & \textbf{0.475} \\  \hline
& & \multicolumn{5}{c}{\textbf{TREC CAsT baselines}}\\\hline
clacBase~\cite{Clarke2019WaterlooClarkeAT}         & -    & -     & -     & 0.246 & 0.640 & 0.360 \\
HistoricalQE~\cite{QueryAnswerExpansionfromConversationHistory} & BERT & -     & -     & 0.267 & 0.715 & 0.436 \\ \hline
& & \multicolumn{5}{c}{\textbf{Manual baselines}}\\
\hline
 Manual           & - & 0.820 & 0.590 & 0.327 & 0.694 & 0.406 \\
Manual           & BERT & 0.820 & 0.757 & 0.389 & 0.857 & 0.577 \\ 
\bottomrule
\end{tabular}
\end{table}

\subsubsection{Conversational Answer Generation.}
\begin{figure}[t]
    \centering
    \includegraphics[width=0.48\textwidth]{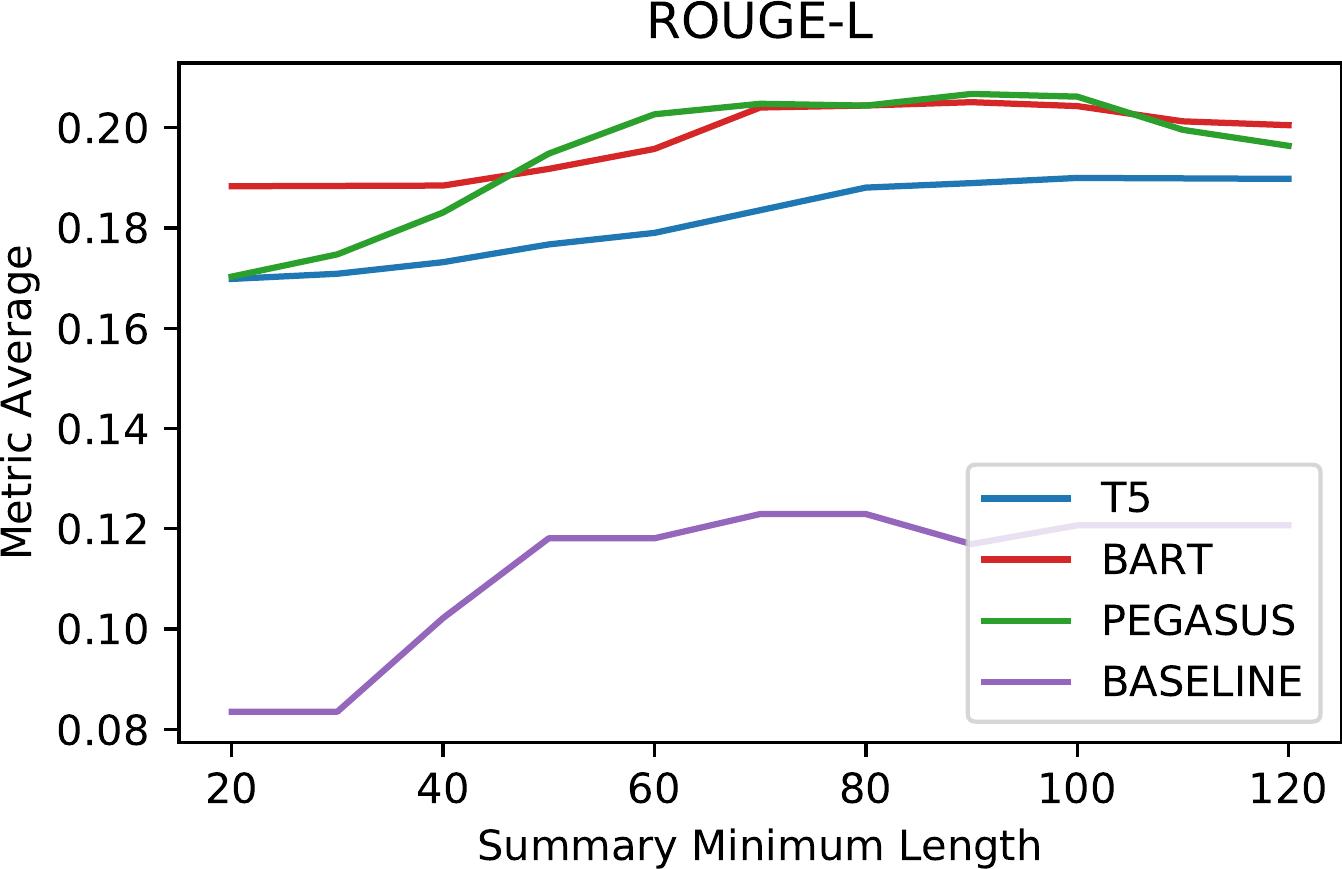}\hspace{3mm}
    \includegraphics[width=0.48\textwidth]{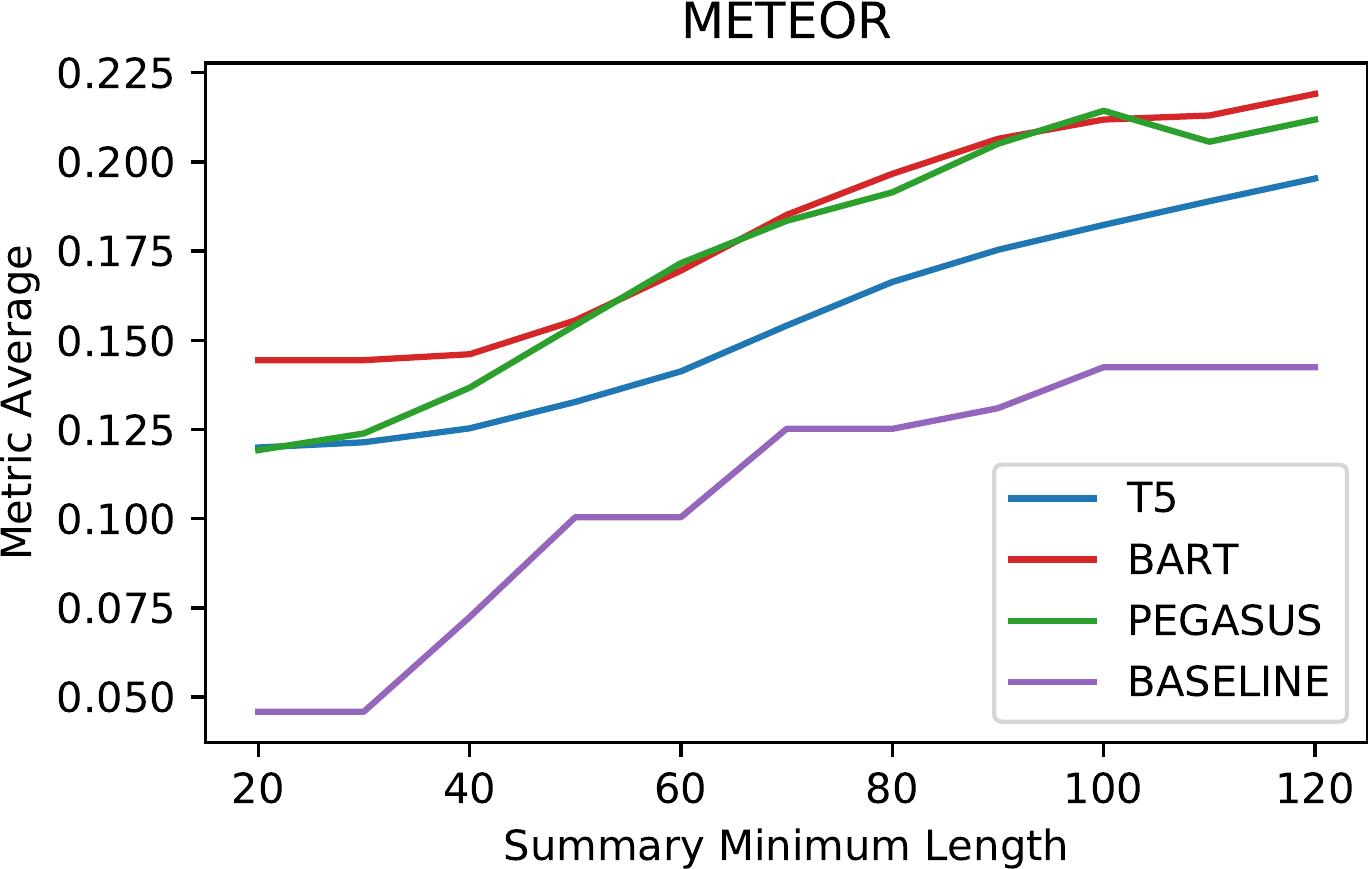}
    \caption{Performance of the answer generation results under different metrics.}
    \label{fig:answerGen}
\end{figure}

Figure~\ref{fig:answerGen} shows the result of the answer generation step according to the ROUGE-L and METEOR metrics. 
The baseline is composed by the concatenation of the top 3 passages, cropped to the maximum length of the passage according to the ``Summary Minimum Length'' value, respecting sentence endings.
In Figure~\ref{fig:answerGen} all answer generation models were better than the retrieval baseline method.
According to ROUGE-L the top performance is achieved around 60-90 word length answers.
Since the goal is to generate short and informative answers, we were not interested in answers longer than 100 words . Actually, we believe that answers with fewer than 50 words are more natural for conversational scenarios. According to these results we observe that BART was the best answer generation method.

\begin{figure}[ht]
    \centering
    \includegraphics[width=0.79\textwidth, trim=0pt 3pt 1pt 0pt, clip]{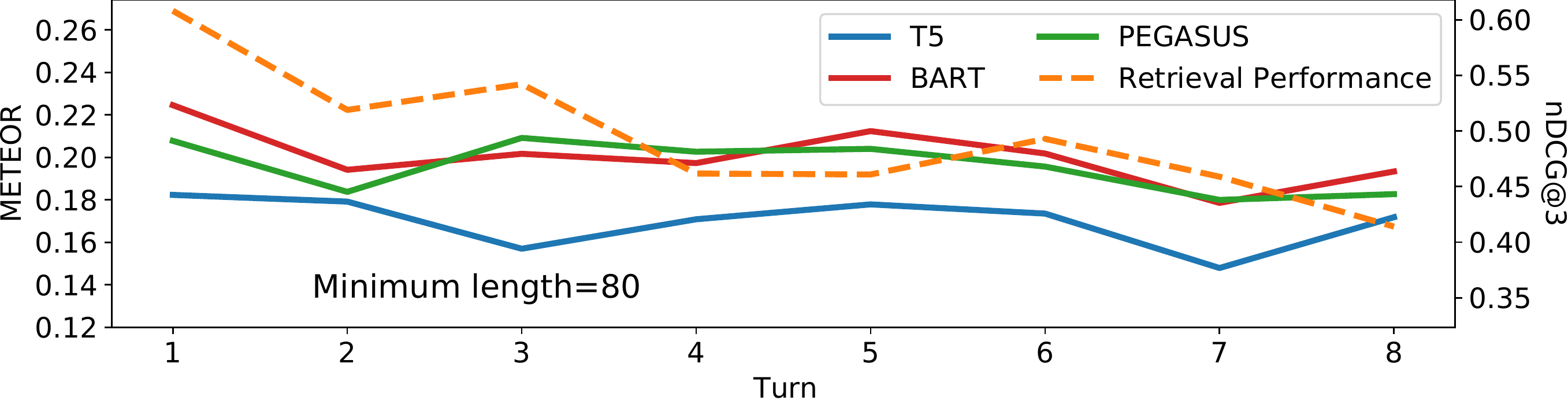}
    \includegraphics[width=0.79\textwidth, trim=0pt 3pt 1pt 0pt, clip]{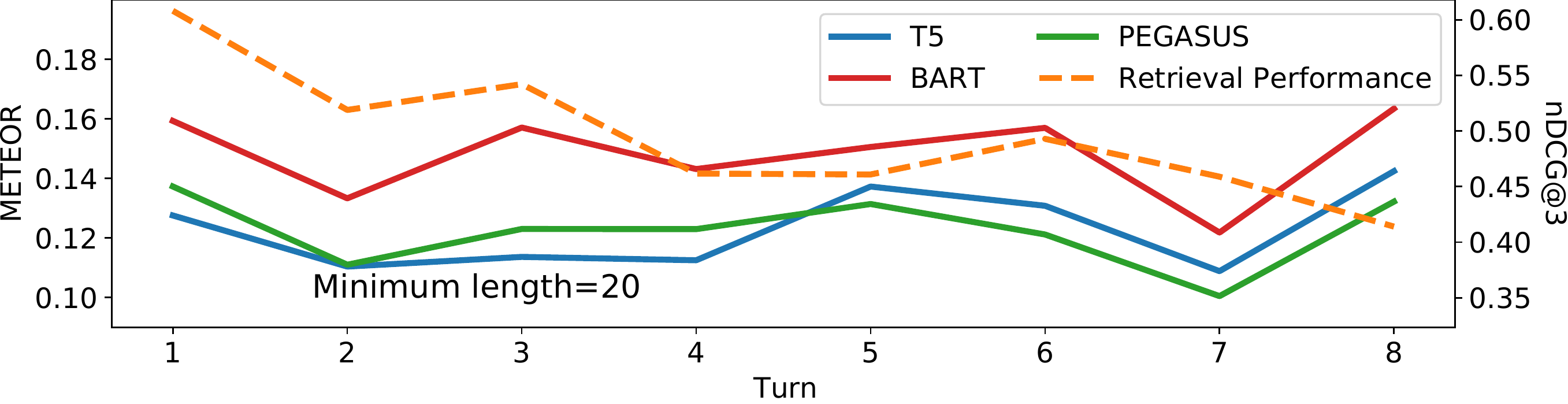}
    \caption{Answer generation versus retrieval performance per conversation turn. The minimum length is 80 and 20 in the top and bottom graphs respectively.}
    \label{fig:answerGenOverTurns}
\end{figure}

In Figure~\ref{fig:answerGenOverTurns} we analyse the retrieval and the answer generation performance over conversation turns. 
We see that peak performance is achieved on the first turn, which was expected given that the first turn that establishes the topic. 
As the conversation progresses, retrieval performance decreases, but surprisingly, answer generation performance is stable until the 6th turn.
We also observed that the decreases in performance are linked to sub-topic shifts within the same conversation topic.

An interesting observation from Figure~\ref{fig:answerGenOverTurns} is that PEGASUS is the method that exhibits a stronger correlation with retrieval performance.
We believe this is related to its generation process that has a behaviour closer to extractive summarisation, while BART and T5 demonstrate a more abstractive behaviour.

Finally, in Table~\ref{table:answer_generation_example} we illustrate the answer generation with all three Transformers.
This table further confirms the abstractive versus extractive summarisation behaviours of the different Transformer-based architectures. 
In this example we see that T5 tries to generate new sentences by combining different sentences.

\begin{table}[ht]
\centering
\caption{Answer generation example for the turn \textit{"What was the first artificial satellite?"}. 
Summary minimum length is set to 90. Blue sentences illustrate abstractive, green sentences illustrate extractive, and red sentences illustrate wrong summaries.}
\label{table:answer_generation_example}
\resizebox{\textwidth}{!}{%
\begin{tabular}{p{0.145\linewidth}p{0.95\linewidth}}
\toprule
\textbf{Method} &  \textbf{Answer} \\ \hline
\textbf{Retrieval Passage 1} &  The first artificial satellite was Sputnik 1, launched by the Soviet Union on October 4, 1957, and initiating the Soviet Sputnik program, with Sergei Korolev as chief designer (there is a crater on the lunar far side which bears his name). This in turn triggered the Space Race between the Soviet Union and the United States. \\ 
\textbf{Retrieval Passage 2} &  The first artificial Earth satellite was Sputnik 1. Put into orbit by the Soviet Union on October 4, 1957, it was equipped with an on-board radio-transmitter that worked on two frequencies: 20.005 and 40.002 MHz. Sputnik 1 was launched as a step in the exploration of space and rocket development. While incredibly important it was not placed in orbit for the purpose of sending data from one point on earth to another. And it was the first artificial satellite in the steps leading to today's satellite communications. \\
\textbf{Retrieval Passage 3} &  The first artificial satellite was Sputnik 1. It was the size of a basketball and was made by the USSR (Union of Soviet Socialist Republics) or Russia. It was launched on October 4, 1957. \\ \hline
\textbf{T5} & \textcolor{ForestGreen}{the first artificial satellite was Sputnik 1, launched by the} \textcolor{red}{u.s. or Russia.} \textcolor{blue}{it was the size of a basketball and launched on October 4, 1957.} \textcolor{ForestGreen}{the satellite} \textcolor{blue}{was equipped with a radio-transmitter that worked on two frequencies.} \textcolor{ForestGreen}{incredibly important it was not placed in orbit for sending data from one point on earth to another. in turn, it triggered the space race between the united states and the soviet union.} \\ \hline
\textbf{BART} & \textcolor{ForestGreen}{The first artificial satellite was Sputnik 1, launched by the Soviet Union on October 4, 1957. It was equipped with an on-board radio-transmitter that worked on two frequencies: 20.005 and 40.002 MHz. This in turn triggered the Space Race between the Soviet Union and the United States.} \textcolor{blue}{The size of a basketball, it was not placed in orbit for the purpose of sending data from one point on earth to another.} \textcolor{ForestGreen}{And it was the first Artificial satellite in the steps leading to today's satellite communications.} \\ \hline
\textbf{PEGASUS} & \textcolor{ForestGreen}{The first artificial satellite was Sputnik 1, launched by the Soviet Union on October 4, 1957. Sputnik 1 was launched as a step in the exploration of space and rocket development. It was not placed in orbit for the purpose of sending data from one point on earth to another. This in turn triggered the Space Race between the} \textcolor{blue}{USSR and the U.S.} \textcolor{red}{There is a crater on the lunar far side which bears his name.} \\
\bottomrule
\end{tabular}%
}
\end{table}
\clearpage
\section{Conclusions}
In this paper we investigated how Transformer architectures can address different tasks in open-domain conversational search, with particular emphasis on the search-answer generation task. The key findings are:
\begin{itemize}
    \item \textbf{Transformers-based Conversational Search.}  Transformers can solve a number of tasks in conversational search, leading to new state-of-the-art results by outperforming the best TREC-CAsT 2019 baseline by $3.9\%$ in terms of nDCG@3. This result is rooted on a fine-tuned bi-directional Transformer model~\cite{t5} for conversational query re-writing, which attained an  improvement of $5-20\%$ (nDCG@3) over raw conversational queries. Similarly, the re-ranking task using a fine-tuned BERT LARGE model~\cite{passagererankingbert} improved results by $10-15\%$ (nDCG@3) over an LMD model.
    \item \textbf{Search-Answer Generation.} Experiments showed that search systems can be improved with agents that abstract the information contained in multiple documents to provide a single and informative search answer. 
    In terms of ROUGE-L we concluded that all answer generation models~\cite{bart,t5,pegasus} performed better than the retrieval baseline.
    \item \textbf{Abstractive vs Extractive Answer Generation.}  The examined answer generation Transformers revealed different behaviours. BART was the most effective in generating answers that were rewritten with information from different passages. This approach turned out to be better than extractive methods that copy and paste sentences from different passages.
\end{itemize}

As future research, we plan to improve conversational query rewriting methods, re-rankers with a notion of the context of the conversation, and mine possible conversation paths to steer the answer generation process towards further helping the user in exploring alternative aspects of the searched topic.

\hfill

\noindent
\textbf{Acknowledgements.} This work has been partially funded by 
the iFetch project, Ref. 45920 co-financed by ERDF, COMPETE 2020, NORTE 2020, 
the CMU Portugal project GoLocal Ref. CMUP-ERI/TIC/0046/2014 and by the project NOVA LINCS Ref. UID/CEC/04516/2013.

\clearpage
\bibliographystyle{splncs04}

\end{document}